\documentclass[preprint]{JASA}

\usepackage{amsmath}
\usepackage{amssymb}
\usepackage{mathtools}
\usepackage{graphicx}
\usepackage{gensymb}
\usepackage{placeins}

\usepackage{makecell}

\begin{document}

\title[]{An infrasound source analysis of the OSIRIS-REx sample return capsule hypersonic re-entry}

\author{Jordan W. Bishop}\email{jwbishop@lanl.gov}\affiliation{Los Alamos National Laboratory, Los Alamos, NM 87545, USA}
\author{Philip Blom}\affiliation{Los Alamos National Laboratory, Los Alamos, NM 87545, USA}
\author{Chris Carr}\affiliation{Los Alamos National Laboratory, Los Alamos, NM 87545, USA}
\author{Jeremy Webster}\affiliation{Los Alamos National Laboratory, Los Alamos, NM 87545, USA}

In Press at the \textit{Journal of the Acoustical Society of America}

\begin{abstract}
\nolinenumbers
The OSIRIS-REx sample return capsule hypersonic re-entry into the atmosphere is a rare opportunity to test a variety of sonic boom source models since the projectile dimensions are well characterized. While the as-flown flight path is unknown, the predicted flight path enables a rough approximation of the source Mach number and location. Six infrasound microphones deployed in the boom carpet along the predicted flight path recorded impulsive signals from the OSIRIS-REx re-entry. Using a suite of atmosphere profiles and the geometric acoustics approximation, we estimate locations with uncertainty estimates along the flight path from which the signals were emitted. Acoustic overpressure and signal duration predictions from Whitham's far field theory, Carlson's simplified sonic boom prediction method, and a drag-dominated hypersonic model are analyzed with uncertainty estimates from the location estimate. While the Carlson simplified sonic boom prediction method could be accurate, our preference is for the drag-dominated source model. Using this source model with an inviscid Burgers' equation solver for propagation, we obtained an excellent match to the recorded data. These results will help better inform future sample return capsule re-entry observation campaigns as well as contribute to a better understanding of high altitude infrasonic sources.
\end{abstract}

\maketitle

\nolinenumbers
LA-UR-25-24888

\section{Introduction}
Infrasonic waves are low-frequency acoustic waves ($\leq$ 20 Hz) that can propagate long distances through the windy, attenuating atmosphere. Infrasound is emitted from many natural sources, such as bolides and lightning, as well as anthropogenic sources such as explosions, industrial activity, and flight vehicles \citep[e.g.,][]{Balachandran1971}. Objects moving through the atmosphere faster than the local sound speed generate a conical shock wave called a sonic boom. The ratio of the object's speed \textit{v} to the local adiabatic sound speed \textit{$c_0$} is termed the Mach number \textit{M}, which is used to characterize different fluid flow regimes around the object \citep{Anderson1984}. The supersonic flow regime occurs when the fluid flow has a Mach number greater than one in the entire region of interest and is characterized by discontinuities in the stream lines of the flow. A fundamental aspect of this regime is that perturbations to the flow can only propagate downstream. At Mach numbers above approximately five, the flow enters the hypersonic regime, where high temperatures and interactions between the shock wave and the viscous boundary layer on the object, among other physical effects, become increasingly important flow phenomena \citep{Anderson1984, Anderson2006}. Sonic booms from supersonic or hypersonic motion can travel large distances (100s to over 1000 kilometers) at infrasonic frequencies through the atmosphere \citep[e.g.,][]{Donn1968, Liszka1978, Rogers1980}, where they eventually pass from the weak-shock into the linear acoustics regime.

Carrying samples from the near-Earth asteroid 101955 Bennu, the NASA (National Aeronautics and Space Administration) OSIRIS-REx (Origins, Spectral, Interpretation, Resource Identification, and Security-Regolith Explorer) SRC (sample return capsule) re-entered Earth's atmosphere on September 24th, 2023 \citep{Lauretta2017}. The SRC obtained hypersonic speeds during its trajectory, which entered Earth's atmosphere over California before touching down at the Utah Test and Training Range in Dougway, Utah \cite{Aljuni2015, Lauretta2017, Francis2024}. The resulting geophysical signals (acoustic, seismic, and electromagnetic) were recorded by a collaborative, multi-institution effort \citep{Silber_2024_OREX}. Joining Genesis (2004), Stardust (2006), Hayabusa 1 (2010), and Hayabusa 2 (2020), the OSIRIS-REx SRC was just the fifth SRC to be recorded acoustically as it returned to Earth since the end of the Apollo missions in the 1970s \citep{Silber2023}. Given the well-known SRC dimensions, the predicted atmospheric re-entry speed of 12 km/s \citep{Lauretta2017}, and the relatively well-characterized trajectory, this re-entry allows a rare test of acoustic source and propagation models for a hypersonic ``artificial meteor" \citep{Ceplecha1998, Silber2023}.

The physics of sonic boom generation and propagation from supersonic aircraft is well known, with many studies examining the effects of sonic booms on structures as well as human annoyance towards civilian supersonic transport \citep[e.g.,][]{Carlson1972, Plotkin2002, Maglieri2014}. Sonic booms from supersonic aircraft, such as the Concorde, have also been proposed as a source for atmospheric sounding, since flight paths can be well parameterized and documented, which is relatively rare for infrasonic sources \citep{Balachandran1971, LePichon2002}. While supersonic aircraft are confined to the troposphere or lower stratosphere, hypersonic acoustic sources are often associated with objects entering or leaving Earth's atmosphere, such as rocket launches, bolides, and the re-entries of SRCs and space shuttles \citep{Ceplecha1998, Henneton2015, Nemec2017, Blom2017, Pilger2021}. Thus, recorded infrasound from these sources may have propagated to the ground from stratospheric to thermospheric altitudes \citep{Maglieri2011_Shuttle}. Characterization of the OSIRIS-REx infrasound would provide additional insight to the emission of infrasound from high altitude acoustic sources as well as better inform instrument campaigns for future SRC re-entries.

The state-of-the-art in sonic boom modeling features direct numerical simulation of the near-source flow field before propagation to farther distances with geometric acoustics \citep[e.g.,][]{Loubeau2009, Luquet2015, Henneton2015, Nemec2017}. These numerical models can leverage a real gas equation of state, implement viscous boundary layer effects, and capture realistic source geometry among other capabilities. These considerations result in more accurate waveform predictions but incur significant computational cost. Our goal in this work is to evaluate the relative effectiveness of simplified, reduced-order models with minimal computational expense. 

This manuscript is organized as follows; we first describe infrasound recordings from the primary boom carpet of the OSIRIS-REx SRC re-entry. We then analyze local atmospheric variability and estimate an approximate source location along the trajectory with uncertainty estimates. A source and propagation analysis is then conducted by evaluating predictions from Whitham's far field theory \citep{Whitham1952}, Carlson's simplified sonic boom prediction method \citep{Carlson1978}, and Tiegerman's drag-dominated hypersonic source model \citep{Tiegerman1975}. Finally, we compare predicted waveforms, with uncertainty due to the propagation, with the recorded data.

\begin{figure}
    \centering
    \includegraphics[scale=1.3]{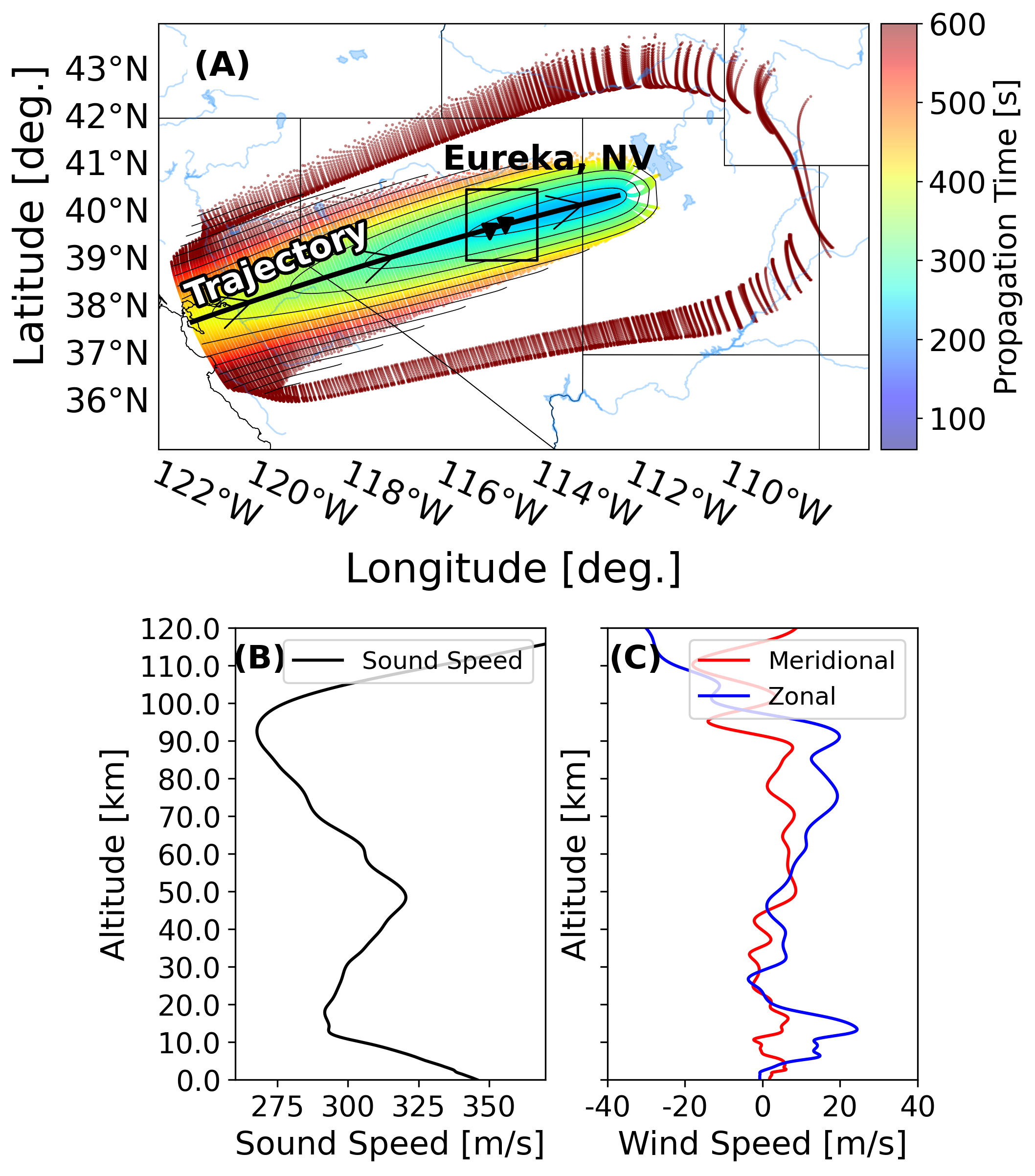}
    \caption{(A) Location of Eureka, NV with deployed microphone locations (black triangles), the post-flight simulated trajectory (thick black line; \citet{Francis2024}), and the ensonified region predicted from \textit{InfraGA} \citep{Blom2024_MachCone}. The arrows along the trajectory denote the direction of motion. Ray piercing points are shaded by the predicted propagation time, and the thin black lines denote isochrones. The elliptical isochronic regions indicate that the source is descending through the atmosphere. Ray paths with turning heights above 120.0 km, which would form additional boom carpets, are not shown. This simulation used the Naval Research Laboratory Ground-to-Space atmosphere profile at Eureka, NV on 09/24/2023 at 15:00:00 UTC. The adiabatic sound speed and wind speeds (meridional and zonal) are shown in (B) and (C), respectively.}
    \label{fig:trajectory}
\end{figure}

\section{Infrasound Data}

\begin{figure}
    \centering
    \includegraphics[scale=1.0]{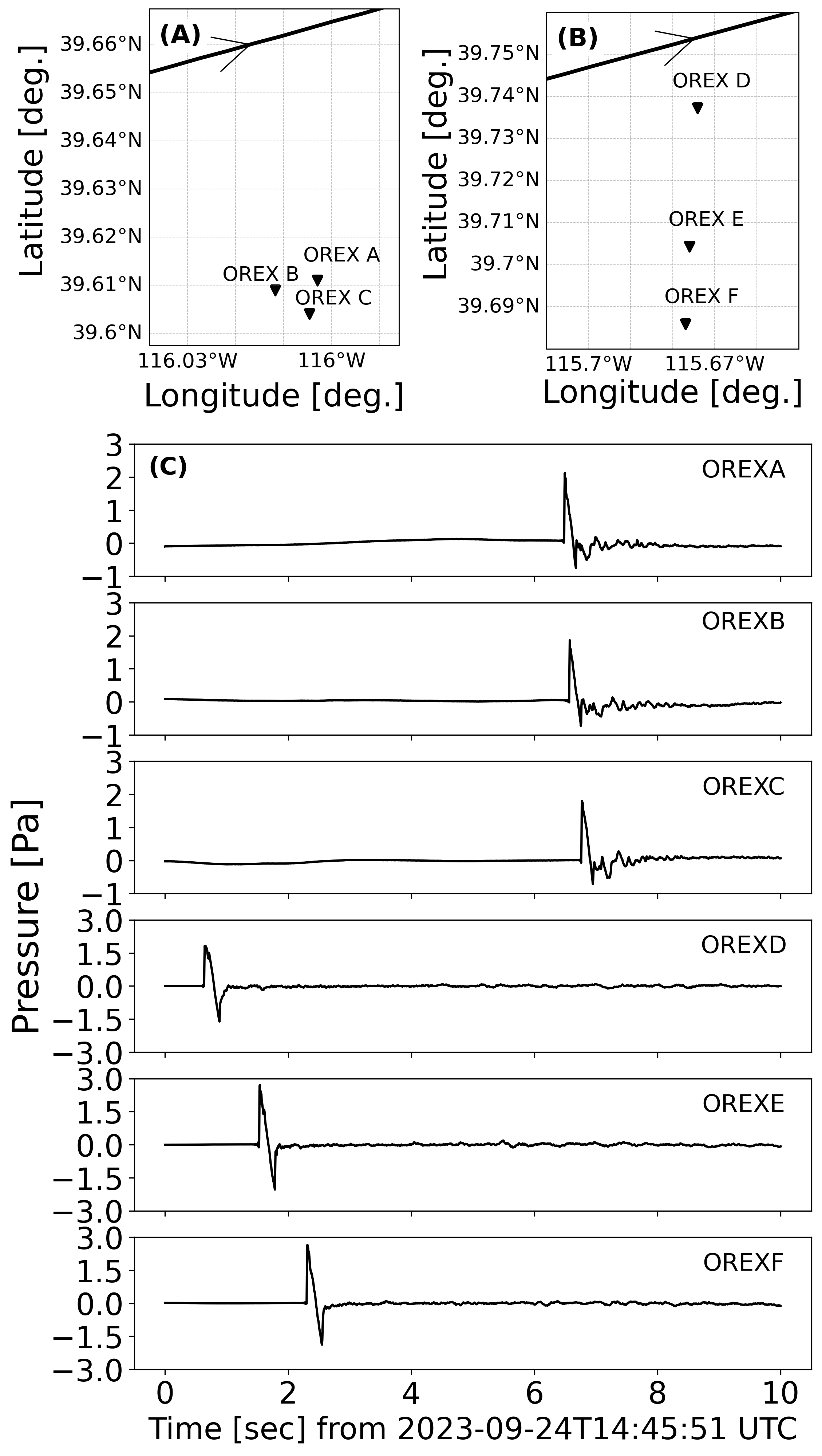}
    \caption{Microphone locations (black triangles) with recorded waveforms. (A) The Eureka Airport microphone geometry south of the simulated, post-flight trajectory (black line; \citet{Francis2024}). (B) The Newark Valley infrasound microphones south of the predicted SRC trajectory. (C) The unfiltered, recorded waveforms at OREXA - OREXF. Note that the y-axis scale bar is different between the two locations.}
    \label{fig:stns}
\end{figure}


In the three days preceding the September 24th, 2023 OSIRIS-REx SRC re-entry \citep{Lauretta2017}, a Los Alamos National Laboratory team deployed infrasound microphones, seismometers, and two distributed acoustic sensing (DAS) interrogators near Eureka, NV (Figure \ref{fig:trajectory}) as part of a broader, multi-institutional effort \citep{Silber_2024_OREX}. The infrasound microphones and seismometers were installed to support interpretation of the DAS recordings, which are detailed in \citet{Carr2025}. The six infrasound microphones are Hyperion 3000 models with a sample rate of 200 Hz, and windscreens are installed over the microphones as a wind noise reduction system \citep{Carr2025}. The deployments were in two clusters south of trajectory, at Eureka Airport, NV (Figure \ref{fig:stns}A) and in Newark Valley, NV (Figure \ref{fig:stns}B).

\begin{table}
\begin{center}
    \caption{Location of infrasound microphones with recorded overpressure $\Delta p$, positive phase duration $\Delta t$, signal time duration between the absolute maximum and the absolute minimum $\tau$, and UTC arrival time on 2023-09-24.}
    \label{tbl:mics}
    \begin{tabular}{ c c c c c c c c}
    Name & Latitude [deg.] & Longitude [deg.] & Elevation [m] & $\Delta p$ [Pa] & $\Delta t$ [s] & $\tau$ [s] & Arrival Time [UTC] \\
    \hline
     OREXA & 39.6109883 & -116.002932 & 1812 & 2.05 & 0.145 & 0.195 & 14:45:57.480 \\ 
     OREXB & 39.60899 & -116.011737 & 1812 & 1.85 & 0.138 & 0.190 & 14:45:57.565 \\  
     OREXC & 39.6040433 & -116.004643 & 1814 & 1.80 & 0.147 & 0.192 & 14:45:57.758 \\ 
     OREXD & 39.7372017 & -115.674093 & 1825 & 1.80 & 0.160 & 0.250 & 14:45:51.635 \\ 
     OREXE & 39.7043 & -115.676033 & 1830 & 2.70 & 0.155 & 0.260 & 14:45:52.525 \\  
     OREXF & 39.6858783 & -115.676975 & 1848 & 2.60 & 0.155 & 0.250 & 14:45:53.200
    \end{tabular}
\end{center}
\end{table}

The recorded waveforms from Eureka Airport (stations OREXA, OREXB, and OREXC) and Newark Valley (stations OREXD, OREXE, OREXF) are shown in Figure \ref{fig:stns}C. All waveforms show an impulsive, N-wave arrival with overpressure $\Delta p$ values between 1.80 and 2.70 Pa (Table \ref{tbl:mics}). The positive phase duration $\Delta t$ values range between 0.138 and 0.160 seconds (Table \ref{tbl:mics}). The waveforms are unfiltered to avoid filter artifacts from the impulsive signal arrival.
As noted in aircraft sonic boom studies \citep{Carlson1972}, signal durations longer than approximately 0.1 seconds may be heard as a ``double boom", which was reported by four observers in Newark Valley \citep{Silber_2024_OREX}. The Eureka Airport data appear to show a signal from a turbulent wake after the main N-wave, but this signal is much weaker, if present, at the Newark Valley microphones 32 km to the northeast. The Newark Valley data appear to have larger amplitudes and positive phase durations compared to the Eureka Airport data (Table \ref{tbl:mics}), and these signals arrive before the Eureka Airport signals since the speed of the descending SRC is faster than the average sound speed along the acoustic propagation path (Figures \ref{fig:trajectory} and \ref{fig:stns}C). 

\begin{figure}
    \centering
    \includegraphics[scale=0.5]{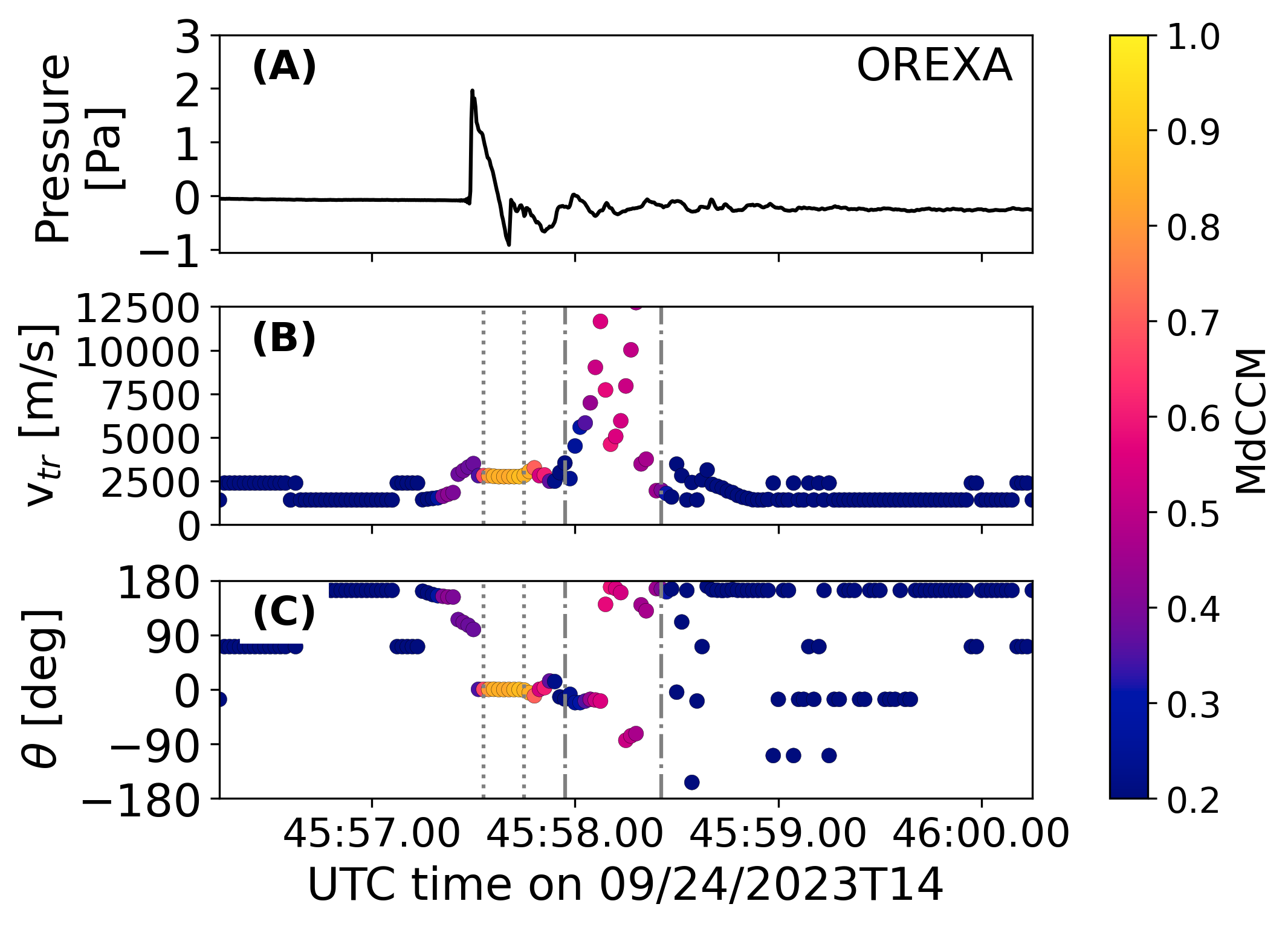}
    \caption{Array processing results from stations OREXA, OREXB, and OREXC (Figure \ref{fig:stns}A). (A) The unfiltered waveform from OREXA, which is replicated here from Figure \ref{fig:stns}C. (B) Estimated trace velocity v$_{tr}$ and (C) back azimuth $\theta$. Warmer colors denote a higher median of the cross correlation maxima (MdCCM) with cooler colors denoting a lower median cross correlation maxima value. The dotted region denotes the shock wave and the dash-dotted region denotes the turbulent wake signal. We note that there is a 0.25 second lag (half the window length) in the trace velocity and back azimuth times due to the use of overlapping, discrete time windows in the array processing.}
    \label{fig:fig2}
\end{figure}

Treating the group of three Eureka Airport microphones (OREXA, OREXB, and OREXC) as an approximately 699 m aperture array, we use a time-domain least squares inversion method \citep{Bishop2020} to estimate trace velocity v$_{tr}$, also called an apparent velocity, and back azimuth values $\theta$ for the N-wave signal (Figure \ref{fig:fig2}). The trace velocity is the speed of the signal in the plane of the array, and the back azimuth is the signal direction of arrival clockwise relative to the north. The median of the cross correlation values (MdCCM) is used as a heuristic measure of signal quality, with warmer colors denoting higher signal coherence. To avoid smoothing the impulsive, high signal-to-noise ratio N-wave signal, we process the unfiltered infrasound data in 0.5 second windows with 95\% overlap. Notably, the N-wave signal (dotted line in Figures \ref{fig:fig2}B and \ref{fig:fig2}C) and the turbulent wake region (dot-dashed line in Figures \ref{fig:fig2}B and \ref{fig:fig2}C) have distinct plane wave parameters. These regions were identified as two distint regions with elevated MdCCM values. The N-wave signal parameters (denoted by the warmer colors in Figures \ref{fig:fig2}B and \ref{fig:fig2}C), have a mean trace velocity of 2769 $\pm$ 22 m/s and a mean back azimuth of 0.2$^\circ \pm 0.5^\circ$. These uncertainty values express one standard deviation, which were estimated by first determining the standard deviations $\sigma_{s_x}$ and $\sigma_{s_y}$ of the elements of the optimal slowness vector with components $s_x$ and $s_y$ \citep{Szuberla2004}, and then using a propagation of errors approach to estimate standard deviations for the trace velocity $\sigma_{v_{tr}}$ and back azimuth $\sigma_\theta$ \citep{DeAngelis2020}.
\begin{align}
\sigma^2_{v_{tr}} &\approx \sigma^2_{s_x} s^2_x v_{tr}^6 + \sigma^2_{s_y} s^2_y v_{tr}^6. \\
\sigma^2_\theta &\approx \sigma^2_{s_x} s^2_y v_{tr}^4 + \sigma^2_{s_y} s^2_x v_{tr}^4. \nonumber
\end{align}
We note the the turbulent wake signal is less coherent across the array, which is noted by the cooler colors in Figures \ref{fig:fig2}B and \ref{fig:fig2}C than the initial N-wave signal, and it varies broadly in trace velocity and back azimuth.

The large signal trace velocities in Figure \ref{fig:fig2} are indicative of an elevated source. We estimate the inclination angle relative to the horizontal $\phi$ of the impinging N-wave using the trace velocity estimate and the local speed of sound $c$ at the array, $\phi = \arccos(c/v_{tr})$. The sound speed at Eureka Airport was calculated to be 337 m/s at 14:53 UTC \citep{Fernando2024}. Using the estimated trace velocity of 2769 $\pm$ 22 m/s, we approximate an incidence angle of 83$^\circ$ relative to the horizontal with an uncertainty of less than one degree. The standard deviation for the incidence angle $\sigma_\phi$ is estimated with a propagation of errors approach \citep{Blom2024b}.
\begin{equation}
\label{eqn:incid}
\sigma_\phi^2 = \frac{\sigma_c^2 + \frac{c^2}{v^2_{tr}}\sigma^2_{v_{tr}}}{v_{tr}^2 - c^2}.
\end{equation}
Here $\sigma_c^2$ is the variance in the sound speed estimate at the array. We estimate $\sigma_\phi $ with $c = 337$ m/s, a somewhat arbitrary $\sigma_c$ = 5 m/s, $v_{tr}$ = 2769 m/s, and $\sigma_{v_{tr}}$ = 22 m/s. Varying the uncertainty in the sound speed from 0 m/s to 10 m/s results in incidence angle uncertainty values less than 0.3$^\circ$. The small uncertainty in the incidence angle is driven by the vast difference in trace velocity, caused by the steep incidence angle, and the local sound speed (Equation \ref{eqn:incid}).

\section{Source Localization}
As an initial step to modeling the source of the OSIRIS-REx N-waves, we estimate the location along the simulated, post-flight trajectory \citep{Francis2024} that the initial signal was emitted from and generate ensembles of atmosphere data to better characterize the uncertainty in the acoustic source location.

\subsection{Atmosphere Ensemble Analysis}

\begin{figure}
    \centering
    \includegraphics[width=12cm]{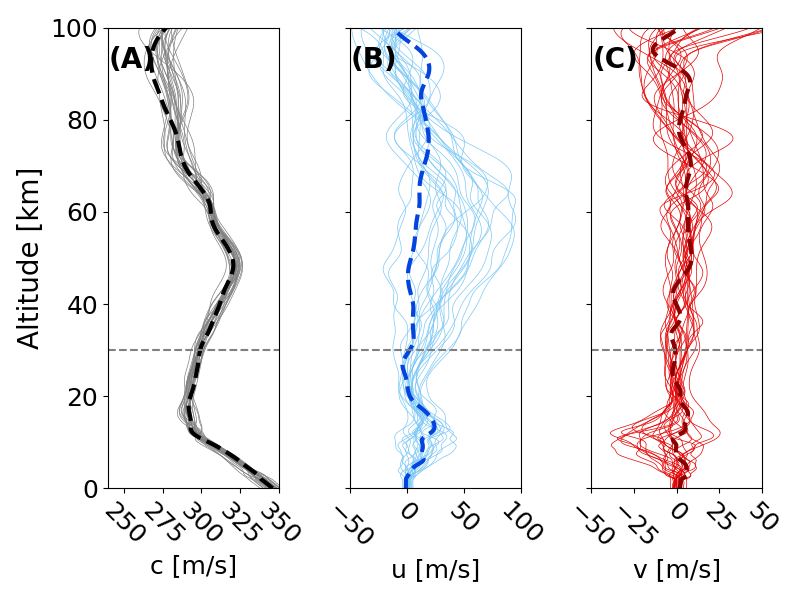}
    \caption{Ensembles of atmosphere sound speed and wind speed profiles for Eureka, Nevada. Panels (A) - (C) depict 100 realizations of (A) the adiabatic sound speed, (B) the zonal wind speed, and (C) the meridional wind speed. The dark, dashed lines in (A) - (C) show the profiles in Figures \ref{fig:trajectory}B and \ref{fig:trajectory}C for reference, and the horizontal dashed line denotes the altitude (approximately 30 km) when the modeled SRC speed drops below Mach 1.}
    \label{fig:atmos}
\end{figure}

Naval Research Laboratory Ground-to-Space atmospheric data from 2023-09-01 to 2023-10-30 were collected at three hour increments for Eureka, Nevada (39.51170$^\circ$ N, 115.96200$^\circ$ W) using the NCPA (National Center for Physical Acoustics) command line tool \citep{Drob2010, Hetzer2024}. 
The atmospheric data were then used to generate empirical orthogonal functions \citep{Blom2023}, which were linearly combined with coefficients drawn from a kernel density function fit to the original set of profiles to generate 100 realizations for the adiabatic sound speed $c$, the zonal wind speed $u$, and the meridional wind speed $v$ (Figures \ref{fig:atmos}A, \ref{fig:atmos}B, and \ref{fig:atmos}C). The ensembles show that, due to including profiles from mid-to-late October 2023, the atmosphere may have both stratospheric and tropospheric winds.

\subsection{Localization}

The conical shock wave emitted by a projectile moving at supersonic or hypersonic speeds is called a Mach cone, which is parameterized by the Mach angle, the half-angle $\alpha = \arcsin(1/M)$ from the axis along the projectile's direction of motion. The Mach cone emits acoustic rays normal to its surface, and the geometric acoustics approximation can be used to efficiently model the propagation of weakly nonlinear shock waves through a layered atmosphere \citep{Plotkin2002}. The earliest algorithms for sonic boom propagation were the ARAP program (Aeronautical Research Association of Princeton; \citet{Hayes1969}) and the so called Thomas algorithm \citep{Thomas1972}. These methods were further developed into codes such as TRAPS (Tracing Rays and Aging Pressure Signatures; \citet{Taylor1980}) and the PCBoom software suite \citep[e.g.,][]{Plotkin1998, Lonzaga2019} among others. In this work, we use the recently developed Mach cone modeling capability of \textit{InfraGA} \citep{Blom2024_MachCone, Blom2019}, a Hamiltonian ray tracer that solves geometric acoustics equations in spherical coordinates (latitude, longitude, altitude).

\begin{figure}
    \centering
    \includegraphics[scale=1.5]{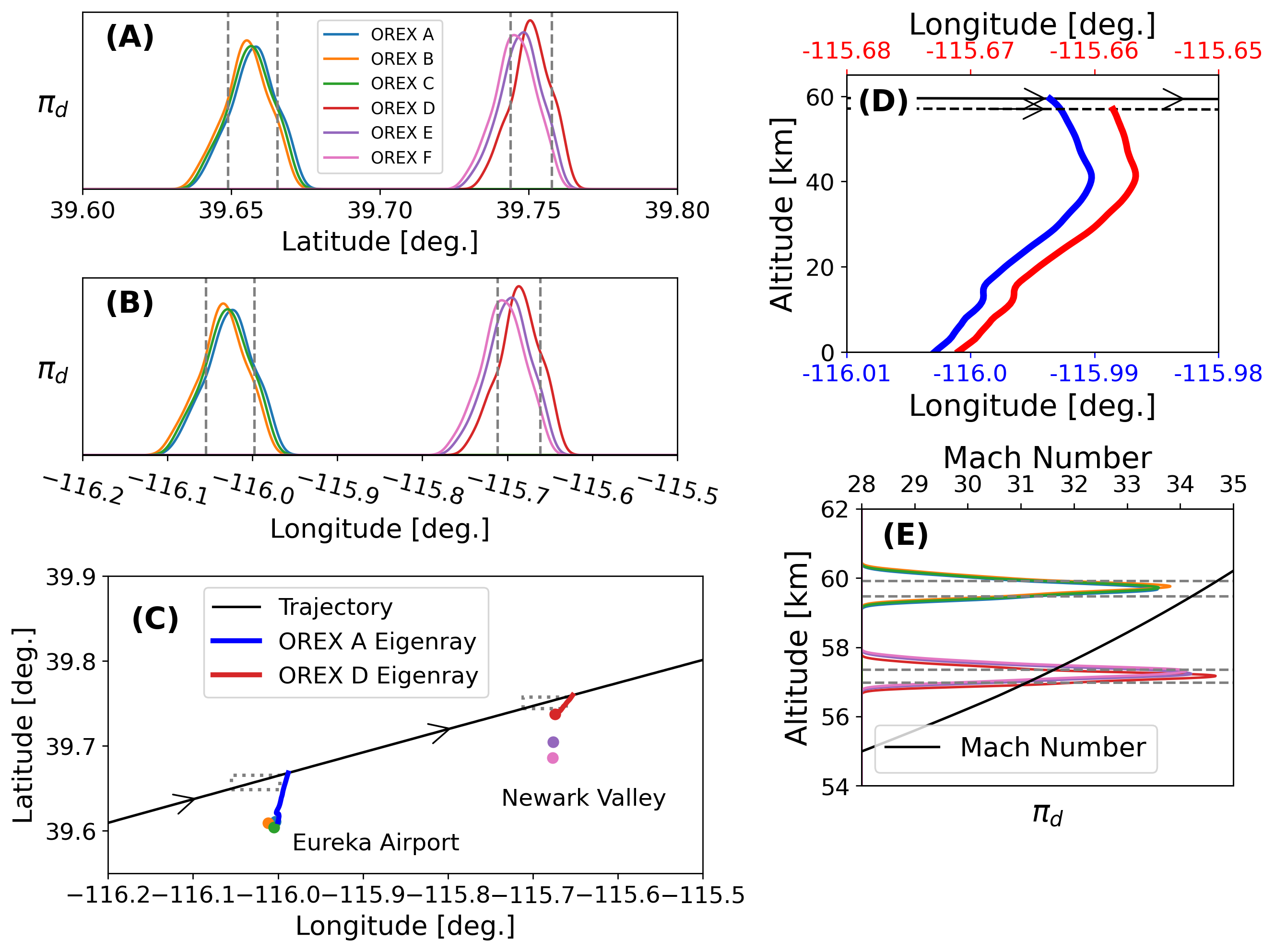}
    \caption{Estimates of N-wave source location for each station. (A) Latitude and (B) longitude probability density estimates $\pi_d$ with dashed gray lines denoting one standard deviation around the mean values for OREXA and OREXD. (C) Map view of the trajectory with station locations (markers) and example eigenrays to OREXA and OREXD. One standard deviation latitude and longitude locations in (A) and (B) are denoted with gray boxes. (D) Vertical transect projected along longitude of the eigenrays in (C). The bottom axis in blue shows the OREXA longitudes, and the top axis in red denotes the OREXD longitudes with the trajectory in this region shown as a dashed line. (E) Altitude probability density estimates $\pi_d$ with the Mach numbers predicted by the post-flight trajectory with one deviation around the OREXA and OREXD mean values. The colors for each sensor are the same across (A) through (E), and the arrows along the trajectories denote the direction of motion.}
    \label{fig:orex_loc}
\end{figure}

We use the nominal EDL (Entry, Descent, and Landing) trajectory \citep{Francis2024} and the 100 atmosphere realizations (Figure \ref{fig:atmos}) to propagate ray paths in a fine grid to the ground (one simulation per atmosphere realization). For each station (Figures \ref{fig:stns}A and \ref{fig:stns}B), we approximate the source location (latitude, longitude, and altitude) as the point along the trajectory that minimizes the distance between the propagated ray pierce point and the station location. Throughout this work, we use kernel density estimation with Gaussian kernels and bandwidth determined by Scott's rule as a non-parametric estimator of underlying probability density functions \citep{Scott1992}. The kernel density estimate fits of the source location ensembles are integrated to calculate the mean source location with one standard deviation uncertainty values, and the results are shown in Table \ref{tbl:Locations} and Figure \ref{fig:orex_loc}.

Example eigenrays to OREXA (blue) and OREXD (red) are shown in Figures \ref{fig:orex_loc}C and \ref{fig:orex_loc}D for reference. The signal source for the Eureka Airport sensors is higher altitude (59.70 $\pm$ 0.22 km for OREXA) and to the southwest of the signal source for Newark Valley (57.17 $\pm$ 0.19 km for OREXD). The location estimates suggest that the source location is approximately north of the sensors at each location, which is consistent with the array processing estimates in Figure \ref{fig:fig2}. Due to the large Mach numbers (Figure \ref{fig:orex_loc}D), the Mach angle is estimated to be less than $2^\circ$.

Using the array estimates from the Eureka Airport sensors (Figure \ref{fig:fig2}), we estimate the source location using a trigonometric approach for comparison \citep[e.g.,][]{Revelle2007}. The ground track of the simulated, post-flight trajectory is approximately 6 km to the north \citep{Francis2024}. With an $83^\circ$ inclination, this suggests a source altitude of approximately 50 km, which is nearly 10 km lower than the estimate with trajectory information (Table \ref{tbl:Locations}). This approach assumes direct signal propagation from the source to the array, such as in a homogeneous atmosphere. The discrepancy may be due to the interaction of ray paths with high altitude winds or the upward refraction of rays near the ground, which would lessen the steepness of an impinging signal and lead to a lower altitude estimate. Location efforts based on travel times \citep[e.g.][]{Nishikawa2022} or back projection \citep{Lonzaga2016, Blom2024b} may lead to increased accuracy and will be pursued in future work.

\begin{table}
\centering
    \caption{Source location estimate of N-waves arriving at each microphone with one standard deviation.}
    \label{tbl:Locations}
    \begin{tabular}{c c c c c}
    {\thead{Element Name}} & {\thead{Source Latitude  [deg.]}} & {\thead{Source Longitude [deg.]}} & {\thead{Source Altitude [km]}} & {\thead{Source Mach Number}} \\
    \hline
     OREXA & 39.657471 $\pm$ 0.007872 & -116.026343 $\pm$ 0.028494 & 59.70 $\pm$ 0.22 & 34.47 $\pm$ 0.23 \\ 
     OREXB & 39.655060 $\pm$ 0.007775 & -116.035119 $\pm$ 0.028145 & 59.77 $\pm$ 0.21 & 34.54 $\pm$ 0.23 \\
     OREXC & 39.656364 $\pm$ 0.007788 & -116.030360 $\pm$ 0.028193 & 59.73 $\pm$ 0.21 & 34.50 $\pm$ 0.23 \\
     OREXD & 39.750608 $\pm$ 0.006876 & -115.686365 $\pm$ 0.025273 & 57.17 $\pm$ 0.19 & 31.38 $\pm$ 0.25\\
     OREXE & 39.747246 $\pm$ 0.007100 & -115.698707 $\pm$ 0.026059 & 57.26 $\pm$ 0.19 & 31.50 $\pm$ 0.26\\
     OREXF & 39.745410 $\pm$ 0.007150 & -115.705468 $\pm$ 0.026252 & 57.31 $\pm$ 0.19 & 31.57 $\pm$ 0.26
    \end{tabular}
\end{table}

\section{Source Modeling}
The state of the art in sonic boom modeling for hypersonic sources couples computational fluid dynamics simulations in the near field with a nonlinear ray method for propagation to longer distances \citep[e.g.,][]{Loubeau2009, Luquet2015, Henneton2015, Nemec2017}. While these source models are superior to analytic models for modeling shock waves from known sources, we opt for parametric, analytical models in this work that lower computational expense and may be of use for future inverse modeling efforts. We evaluate predictions from Whitham's far field theory \citep{Whitham1952}, Carlson's simplified sonic boom prediction method \citep{Carlson1978}, and Tiegerman's drag-dominated hypersonic source model \citep{Tiegerman1975}. For each method, we estimate a reference distance such that the predicted ratio of overpressure to reference pressure $\Delta p / p_\text{ref}$ falls to 0.1. This value is somewhat arbitrary, but it falls within the weak shock regime \citep{Jones1968}. We construct a symmetric N-wave as the initial source for the inviscid Burger's equation using the predicted overpressure $\Delta p_0$ and positive phase duration $\Delta t_0$ and propagate this N-wave to the receivers with an inviscid Burgers' equation \citep{Lonzaga2015, Lonzaga2019, Blom2021_Thermosphere}. Additionally, the Whitham and Carlson theories can be used to directly predict the signal pressure and duration on the ground, and we will evaluate these estimates in addition to the propagated waveforms to evaluate the relative contribution of the weakly nonlinear propagation.

The inviscid Burgers' equation describes the evolution of the scaled acoustic pressure $u$ along rath paths in the weak shock limit
\begin{equation}
\label{eqn:burgers}
\frac{\partial u}{\partial s} = \tilde{\beta} u \frac{\partial u}{\partial \tau}.
\end{equation}
Here $\tilde{\beta}$ is the scaled coefficient of nonlinearity, $s$ is the ray length, and $\tau$ is retarded time. The scaled coefficient of nonlinearity is expressed as 
\begin{equation}
\label{eqn:burgers2}
\tilde{\beta} = \beta \frac{p_\text{ref}}{\rho c}\frac{\psi}{c_g c_\text{src}}\sqrt{\frac{\rho c^3 \psi}{\rho_\text{ref} c_\text{ref}^3 \psi_\text{ref}}\frac{D_\text{ref} c_{g, \text{ref}}}{D c_g}},
\end{equation}
where $\beta = (\gamma + 1) /2 = 1.2$ is the coefficient of nonlinearity, $\gamma = 1.4$ is the ratio of specific heats, $p_\text{ref}$ is a reference pressure, $\rho$ is the ambient atmospheric density, $c_g$ is the group speed along the ray path, $c_\text{src}$ is the sound speed at the source, $\psi$ is the magnitude of the eikonal vector, and $D$ is the Jacobian of the transformation from Cartesian to ray coordinates \citep{Hayes1972, Lonzaga2019, Blom2021_Thermosphere}. We note that this Jacobian expresses cylindrical as opposed to spherical spreading. The ray coordinates for this work are the ray length, the ray azimuth, and the ray inclination angle \citep{Blom2019}. The group velocity is the speed at which energy propagates in the direction of wave motion, and the eikonal vector points normal to surfaces of constant phase. The subscript ``ref" in Equation \ref{eqn:burgers2} denotes values at a reference distance along the ray path at which the waveform is known. The scaled acoustic pressure $u$ is related to the overpressure $\Delta p$ as 
\begin{equation}
u(s,\tau) = \frac{\Delta p (s, \tau)}{p_\text{ref}}\sqrt{\frac{\rho_\text{ref} c_\text{ref}^3 \psi_\text{ref}}{\rho c^3 \psi}\frac{D c_g}{D_\text{ref} c_{g, \text{ref}}}},
\label{eqn:Burgers_eqn}
\end{equation}
where the reference pressure $p_\text{ref}$ is the maximum absolute overpressure of the waveform \citep{Lonzaga2019, Blom2021_Thermosphere}. Equation \ref{eqn:burgers} is solved along the eigenray from the source to the receiver using a Heun's method numerical scheme for time stepping \citep{Blom2021_Thermosphere}.

\subsection{Whitham's Far Field Theory}
\label{section:Whitham}
Early research into ``ballistic waves" established that the compression and rarefaction lobes of the ``N-wave" were balanced, and that the shock wave decays with radial distance from the source $r$ proportionally to $ r^{-3/4}$ \citep{Landau1945, DuMond1946}. Linearized treatments of supersonic flow yielded approximate results, but a comprehensive theory was not established until curvature was introduced to the characteristic lines in the flow field \citep{Whitham1952}. Whitham's theory established that shocks would form where characteristic lines cross and developed a source function that connects the pressure disturbance to the shape of the projectile, termed the F-function.
\begin{equation}
F(y) = \frac{1}{2\pi}\int_0^y \frac{A''(x)}{\sqrt{y - x}} dx.
\label{eqn:F_function}
\end{equation}
Here $x$ is the horizontal distance along the projectile, and $A''(x)$ denotes the second derivative with respect to distance of the the cross-sectional area of the projectile $A(x)$. The term $y$ describes the shape of the characteristic lines and equals $x - r\sqrt{M^2-1}$ on the projectile surface, where $r$ is the distance from the evaluation point to the horizontal axis. The development of the F-function was fundamental for establishing the strength of sonic booms from supersonic aircraft.

In the far field, the contribution of the F-function to the overpressure is summarized as a shape factor K,
\begin{equation}
K = \sqrt{\int_0^{\eta_0}F(\eta)d\eta},
\label{eqn:K_W}
\end{equation}
where $\eta_0$ is the distance along the horizontal axis that maximizes the integral \citep[e.g.,][]{Gottlieb1988}. Equivalently, this maximizer is a root of the F-function. Assuming the projectile can be approximated as a slender cone leads to an approximate value of 
\begin{equation}
K \approx \delta l^{(3/4)},
\label{eqn:K_delta}
\end{equation}
where $l$ is the length of the projectile, and $\delta$ is the slenderness ratio: the maximum radius of the projectile divided by its length. We also stress that, in general, the F-function is an acoustic source function, and it may be approximated outside of the analytical formulations of Equations \ref{eqn:F_function} or \ref{eqn:K_delta}.

The N-wave overpressure $\Delta p$ and positive phase duration $\Delta t$ are predicted as
\begin{align}
\label{eqn:Whitham_eqns}
\Delta p &= \frac{2^{1/4}\gamma p_\text{ref}}{\sqrt{\gamma + 1}}(M^2 - 1)^{1/8}\frac{K}{r^{3/4}},
\\
\Delta t &= \frac{2^{1/4}\sqrt{\gamma + 1}M}{c_\text{src}(M^2 - 1)^{3/8}}Kr^{1/4}. \nonumber
\end{align}
Here $r$ is the radial distance from the source, $c_\text{src}$ is the adiabatic sound speed at the source altitude, and $p_\text{ref}$ is the reference pressure \citep{Whitham1952}. This theory was originally derived for a uniform atmosphere \citep{Whitham1952}, and the reference pressure $p_\text{src}$ for a real atmosphere can be approximated as the geometric mean of the ambient pressure at the projectile $p_\text{src}$ and the ambient pressure at the sensor $p_G$ \citep{Carlson1966}. The OSIRIS-REx SRC has identical dimensions to the Stardust SRC \citep{Mitcheltree1999, Aljuni2015, Francis2024}, so the maximum effective radius is approximately 0.405 m, the length $l$ is approximately 0.5 m, and the slenderness ratio $\delta$ is 0.8. Therefore, the shape factor $K$ is approximately 0.5, which is in line with a previously reported value by \citet{Fernando2025}.

\begin{figure}
    \centering
    \includegraphics[scale=0.5]{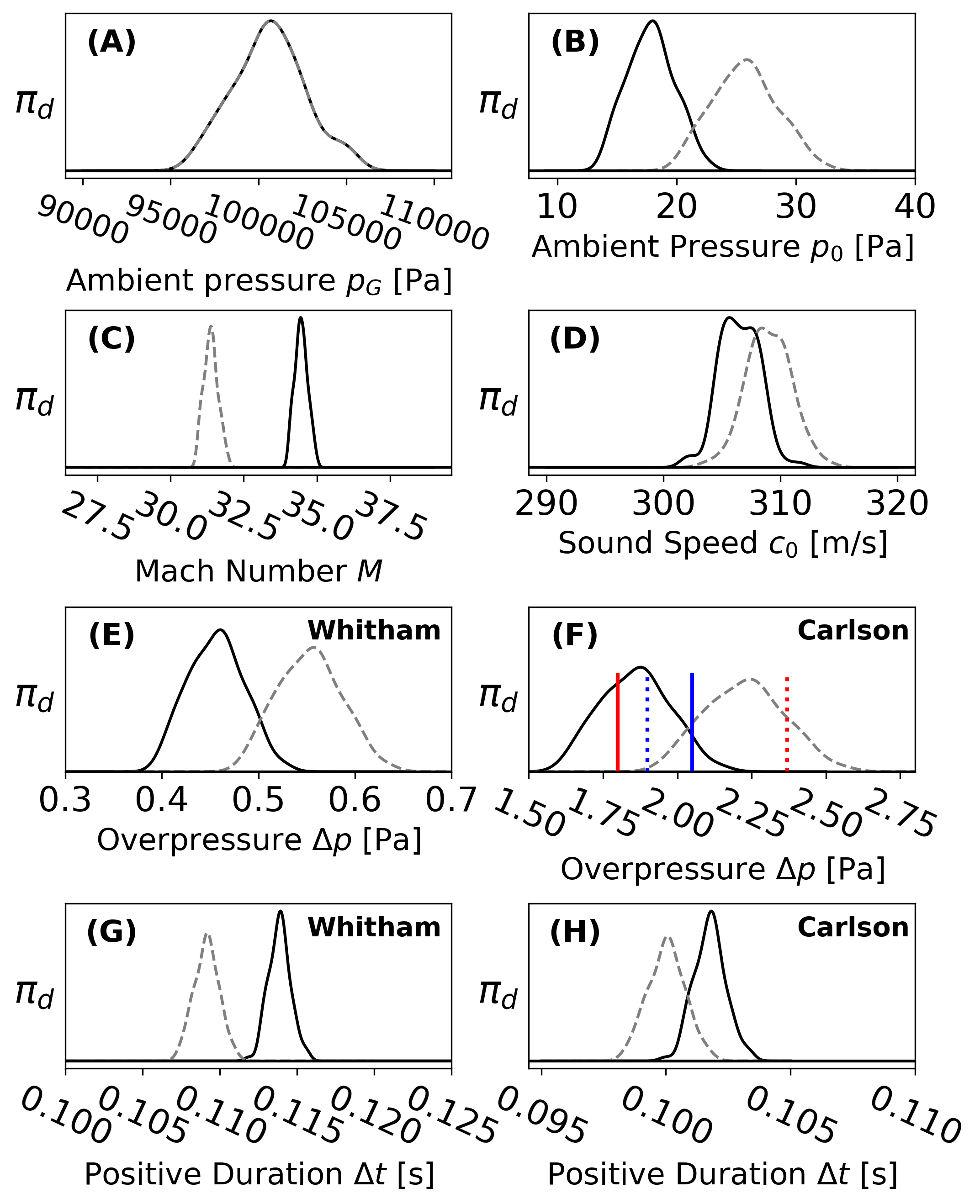}
    \caption{Estimated probability density functions $\pi_d$ of parameters used for predicting overpressure and positive phase duration using Equations \ref{eqn:Whitham_eqns} and \ref{eqn:Carlson_eqns}. Due to the high degree of similarity between stations at Eureka airport and between stations at Newark Valley, estimates for OREXA (black lines) and OREXD (gray dashed lines) are shown as representative values. (A) The ambient pressure at ground level $\rho_G$, which is identical between both locations. (B) The ambient pressure at the estimated source height $\rho_0$. (C) Estimated Mach number from the projected trajectory from NASA. (D) Estimated adiabatic sound speed at the source altitude. Overpressure $\Delta p$ (E) and signal positive phase duration $\Delta t$ (G) predictions using Equations \ref{eqn:Whitham_eqns}. Overpressure $\Delta p$ (F) and signal positive phase duration $\Delta t$ (H) predictions using Equations \ref{eqn:Carlson_eqns}. The recorded N-wave overpressure for OREXA (solid red line)  and OREXD (solid blue line) are shown in (F) with the average values across the Eureka Airport stations (red dotted line) and Newark Valley stations (blue dotted line) for reference (Table \ref{tbl:mics}). Subplots (B), (C), and (D) are consistent with an object slowing down as it descends through the atmosphere.}
    \label{fig:PDFs1}
\end{figure}

Using Equations \ref{eqn:Whitham_eqns}, we predict overpressure and signal durations for N-waves recorded at the Eureka airport and Newark Valley sensors (Figure \ref{fig:stns}). The altitude estimates in Figure \ref{fig:orex_loc}E are substituted for $r$ and are used to calculate corresponding adiabatic sound speed values, ambient pressure values, and Mach numbers (Figure \ref{fig:PDFs1}A - \ref{fig:PDFs1}D). As with the source locations, a kernel density estimate with Gaussian kernels and a bandwidth determined by Scott's rule was fit to the data and used to calculate a mean and standard deviation for the predictions. The probability distribution functions for overpressure and signal duration for OREXA and OREXD are shown in Figures \ref{fig:PDFs1}E and \ref{fig:PDFs1}G. The predicted overpressures are 0.456 $\pm$ 0.030 Pa for OREXA and 0.553 $\pm$ 0.035 Pa for OREXD. The signal positive phase predictions are 0.114 $\pm$ 0.001 s and 0.109 $\pm$ 0.001 s, respectively.

Additionally, Equations \ref{eqn:Whitham_eqns} are used to generate N-waves that are propagated down to the Eureka airport and Newark Valley sensors with the Burgers' equation solver \citep{Blom2021_Thermosphere}. The reference distance was chosen such that the $\Delta p / p_\text{ref}$ ratio equaled $0.1$, and this value was used to generate initial overpressure $\Delta p_0$ and positive phase duration $\Delta t_0$ values for the N-waves. These initial values are shown in Figures \ref{fig:PDFs2}A and \ref{fig:PDFs2}B, the predicted overpressure $\Delta p$ and positive phase duration $\Delta t$ values after propagation to the ground are shown in Figures \ref{fig:PDFs2}C and \ref{fig:PDFs2}D, and the resulting waveforms at the ground are shown in Figure \ref{fig:N_wave_comparison}. The predicted overpressures for this method are 0.971 $\pm$ 0.047 Pa for OREXA and 1.190 $\pm$ 0.055 Pa for OREXD. The corresponding signal positive phase duration predictions are 0.105 $\pm$ 0.003 s (OREXA) and 0.104 $\pm$ 0.002 s (OREXD).

\subsection{Carlson's Simplified Sonic Boom Prediction Method}
\label{section:Carlson}

\begin{figure}
    \centering
    \includegraphics[scale=2.1]{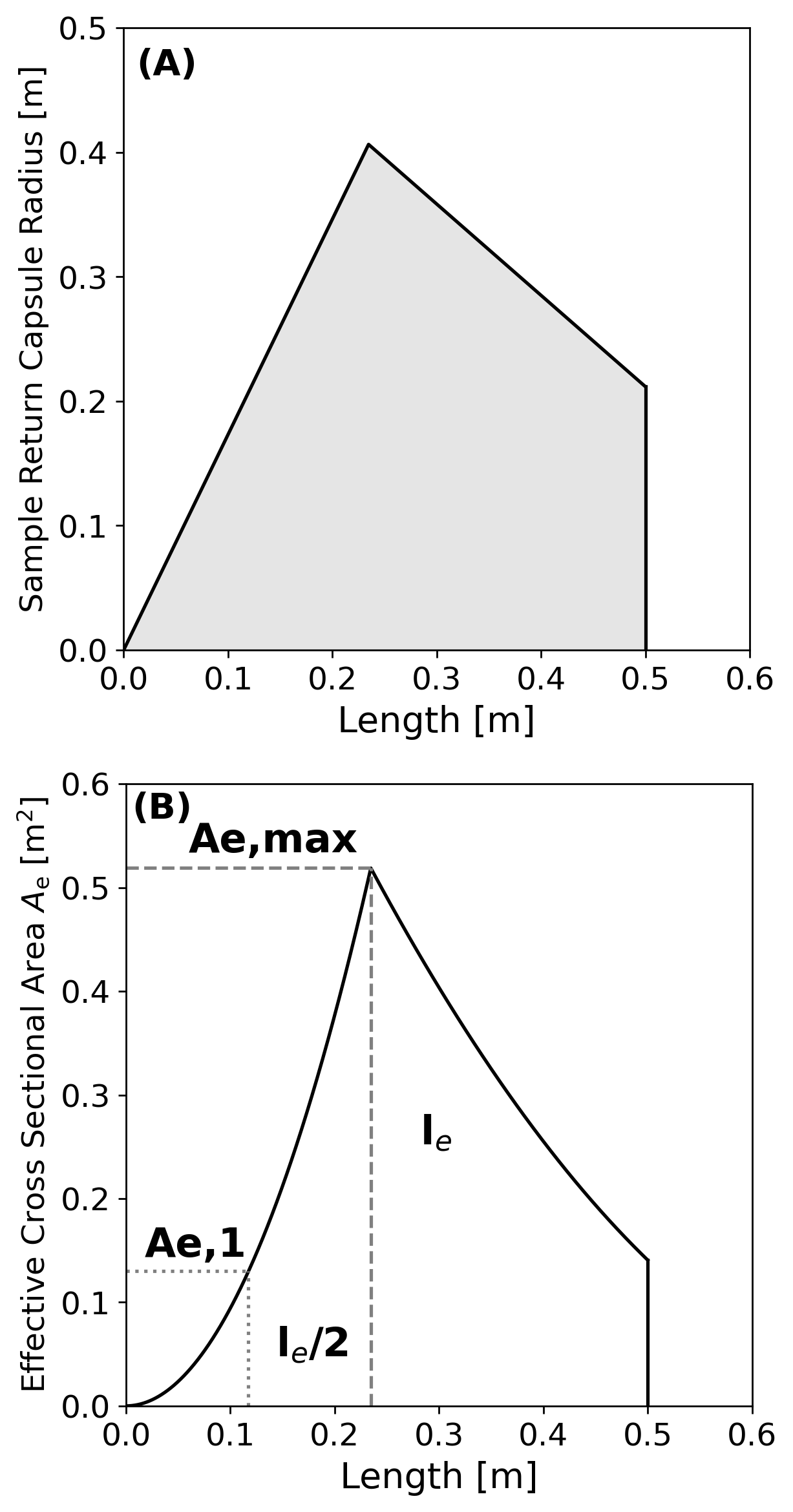}
    \caption{(A) Approximation of the sample return capsule geometry \citep{Mitcheltree1999, Francis2024}. (B) Schematic for calculating the shape factor $K_S$ for Carlson's Simplified Sonic Boom Prediction \citep{Carlson1978}. The cross sectional area of the SRC is shown with the black line, and the effective area values (A$_{e,1}$ and A$_{e_\text{max}}$) shown as a function of the effective length l$_e$ (gray lines).}
    \label{fig:Carlson}
\end{figure}

Carlson's simplified sonic boom prediction method \citep{Carlson1978, Carlson1966} is a graphical method that was developed by adding correction factors for a realistic atmosphere to the Whitham far field theory \citep{Whitham1952}. This method is also a shape factor theory, but the shape factor $K_S$ may be derived graphically as a function of the effective length $l_e$ and effective cross sectional area $A_\text{e}$ of the projectile. The effective cross sectional area also includes a contribution due to lift, which we here consider equal to zero. The shape factor $K_S$ is
\begin{align}
K_S &= C\frac{\sqrt{A_{e,\text{max}}}}{l^{(3/4)}l_e^{(1/4)}}, \\
C &= f\bigg(\frac{A_{e,1}}{A_{e,\text{max}}}\bigg). \nonumber
\end{align}
where $A_{e,\text{max}}$ is the maximum cross sectional area, $l_e$ is the length associated with the maximum cross sectional area, also called the effective length, and A$_{e,1}$ is the effective cross sectional area associated with half the effective length. The factor $C$ is a function of A$_{e,1}$ and $A_{e,\text{max}}$, which is plotted as Figure 2 of \citet{Carlson1978}. Using an approximation to the OSIRIS-REx geometry (Figure \ref{fig:Carlson}A), which was approximated as a 60$^\circ$ cone attached to the frustrum of a 30$^\circ$ cone \citep{Mitcheltree1999}, the effective area is shown in Figure \ref{fig:Carlson}B. Using a length $l = 0.5$ m, the effective length $l_e$ is estimated as 0.23 m, the maximum effective area is estimated as 0.52 m$^2$, and A$_{e,1}$ is estimated as 0.13 m$^2$. We estimate the factor $C = 0.705$, so the shape factor $K_S$ is 1.23.

The prediction equations for overpressure $\Delta p$ and N-wave positive phase duration $\Delta t$ are
\begin{align}
\label{eqn:Carlson_eqns}
\Delta p &= K_p K_R \sqrt{p_0 p_G}(M^2 - 1)^{1/8}r^{-3/4}l^{3/4}K_S,
\\
\Delta t &= K_t \frac{3.42}{2 c_0}\frac{M}{(M^2 - 1)^{3/8}}h_e^{1/4}l^{3/4}K_S. \nonumber
\end{align}
$K_p$ is the pressure amplification factor, $K_R$ is the reflection factor, $p_G$ is the atmospheric pressure at ground level, $h_e$ is the effective altitude, $K_t$ is the signature duration factor, and $c_0$ is the adiabatic sound speed at the source height. The reflection factor $K_R$ is often defined as 2.0, but it may be somewhat less due to finite ground impedance \citep{Coulouvrat2002}. The values for $K_p$ and $K_t$ are provided in graphical form as Tables 7d and 7e in \citet{Carlson1978}. While the trajectory entry angle is predicted to be approximately 8$^\circ$ from the horizontal \citep[e.g.,][]{Silber2023}, given the close location of the recording sensors to the ground track (Figure \ref{fig:trajectory}), we approximate the effective altitude $h_e$ with the estimated source altitude (Figure \ref{fig:orex_loc}B).

We apply the same method as in Section \ref{section:Whitham} to obtain estimates of $p_\text{src}$, $c_\text{src}$, and altitude $h$ for each atmospheric profile in Figure \ref{fig:atmos}. Since this is a graphical method, we use the mean altitude values in Table \ref{tbl:Locations} to estimate constant values of $K_p = 1.485$ and $K_t = 0.66$ for the Eureka Airport stations (OREXA, OREXB, and OREXC), and $K_p = 1.480$ and $K_t = 0.675$ for the Newark Valley stations (OREXD, OREXE, and OREXF). The probability distribution functions for overpressure $\Delta p$ and signal positive phase duration $\Delta t$ are shown in Figures \ref{fig:PDFs1}F and \ref{fig:PDFs1}H, and the mean predictions are summarized in Table \ref{tbl:All_predictions}. The predicted overpressures are 1.859 $\pm$ 0.123 Pa for OREXA and 2.228 $\pm$ 0.140 Pa for OREXD. The signal positive phase predictions are 0.102 $\pm$ 0.001 s and 0.100 $\pm$ 0.001 s, respectively.

Additionally, using the same procedure described Section \ref{section:Whitham}, Equations \ref{eqn:Carlson_eqns} were used to generate N-waves that are propagated down to the Eureka airport and Newark Valley sensors with the Burgers' equation solver. Initial values are shown in Figures \ref{fig:PDFs2}A and \ref{fig:PDFs2}B, the predicted overpressure and signal duration values after propagation to the ground are shown in Figures \ref{fig:PDFs2}C and \ref{fig:PDFs2}D, and the ensemble of waveforms propagated to the ground is shown in Figure \ref{fig:N_wave_comparison}. The predicted overpressures for this method are 1.692 $\pm$ 0.086 Pa for OREXA and 2.108 $\pm$ 0.088 Pa for OREXD. The corresponding signal positive phase predictions are 0.175 $\pm$ 0.010 s (OREXA) and 0.175 $\pm$ 0.004 s (OREXD).

\subsection{Tiegerman 1975 Drag Dominated Hypersonic Model}
\label{section:Tiegerman}
The shock wave produced by steady, two dimensional (axisymmetric) hypersonic flow over a slender, blunt-nosed projectile can equivalently be expressed as unsteady, one dimensional flow produced by a cylindrical point explosion, which is called the \textit{principle of hypersonic similarity} \citep{Hayes1947}. This principle was leveraged by \citet{Tiegerman1975} to develop a far-field hypersonic source model for overpressure $\Delta p$ and positive phase duration $\Delta t$ that replicates $r^{-3/4}$ amplitude decay in the weak shock regime:
\begin{align}
\Delta p &= \frac{0.6079 p_\text{src} R_0}{r} \Big[\sqrt{\frac{r}{R_0}} + 0.55\Big]^{1/2}, 
\label{eqn:Tiegermanp}
\\
\Delta t &= 1.042 T_0 \sqrt{1 - 0.55\sqrt{\frac{R_0}{r}}}\Big(\frac{r}{R_0}\Big)^{1/4};
\label{eqn:Tiegermant}
\end{align}
where $p_\text{src}$ is the ambient pressure at the source, $r$ is the radial distance from the source, $R_0$ is the characteristic length scale, and $T_0$ is the characteristic time scale. As opposed to the Whitham and Carlson shape factor methods described above, the overpressure and positive phase duration values in Equations \ref{eqn:Tiegermanp} and \ref{eqn:Tiegermant} are not meant as a combined source and propagation model. Instead, this theory describes boundary conditions for an inviscid Burgers' equation, which can be used to propagate the signal from the source locations to the microphones.

We assume that the capsule deposits energy into the atmosphere as a result of drag $D$, 
\begin{equation}
D = \frac{1}{2}C_D \rho_\text{src} v^2 \frac{\pi d^2}{4},
\end{equation}
where $C_D$ is the drag coefficient, $\rho_\text{src}$ is the ambient density at the source, $v$ is the speed of the projectile, and $d$ is the diameter of the projectile. The length scale $R_0$ is
\begin{align}
R_0 &= \sqrt{\frac{D}{2 \pi p_\text{src}}}, \nonumber \\
&= \frac{\sqrt{\gamma C_D} M d}{4}.
\label{eqn:R0}
\end{align}
Here $\gamma = 1.4$ is the specific heat ratio of the gas, M is the Mach number relative to the ambient sound speed $c_\text{src}$ at the source, and the substitution $c_\text{src}^2 = \gamma p_\text{src}/\rho_\text{src}$ was also used. The length scale $R_0$ is also called the blast radius or relaxation radius, and it has been defined with different proportionality constants by different authors \citep{Silber2019}. The characteristic time scale $T_0$ is
\begin{equation}
T_0 = \frac{R_0}{c_\text{src}}.
\label{eqn:T0}
\end{equation}
The OSIRIS-REx sample return capsule dimensions are identical to the Stardust sample return capsule dimensions \citep{Mitcheltree1999, Aljuni2015, Francis2024}. From an aerodynamics analysis of the Stardust SRC \citep[][Figure 6]{Mitcheltree1999}, we estimate an axial coefficient $C_A$ of 1.49 for a Mach number of approximately 30. For an angle of attack of $0^\circ$, the drag coefficient $C_D$ is equal to the axial coefficient, so $C_D = 1.49$.

We apply the same method as in Sections \ref{section:Whitham} and \ref{section:Carlson} to obtain estimates of $p_\text{src}$, $c_\text{src}$, and altitude (Figures \ref{fig:orex_loc}B, \ref{fig:PDFs1}B, and \ref{fig:PDFs1}D). Using these values and a SRC diameter $d = 0.81$ meters, we estimate $R_0$ and $T_0$, and then estimate a reference distance $r_\text{ref}$ from Equation \ref{eqn:Tiegermanp} such that the overpressure ratio $\Delta p / p_0$ falls to 0.1. We note that the atmosphere is treated as homogeneous between the source location and the reference distance. The length scale $R_0$ is approximately 10.0 $\pm$ 0.1 m for the signals observed at OREXA and 9.1 $\pm$ 0.1 m for the signals observed at OREXD. The time scale $T_0$ is approximately 0.0328 $\pm$ 0.0003 s for OREXA and 0.0296 $\pm$ 0.0003 s for OREX D. We construct a symmetric N-wave as the initial waveform for the inviscid Burger's equation using the overpressure values $\Delta p_0$ and positive phase duration $\Delta t_0$ values from Equations \ref{eqn:Tiegermanp} and \ref{eqn:Tiegermant}. These source waveforms are then propagated to OREXA and OREXD with the Burgers' equation solver \citep{Lonzaga2015, Lonzaga2019, Blom2021_Thermosphere} without viscous losses \citep{Sutherland2004}. 

\begin{figure}
    \centering
    \includegraphics[width=0.5\textwidth]{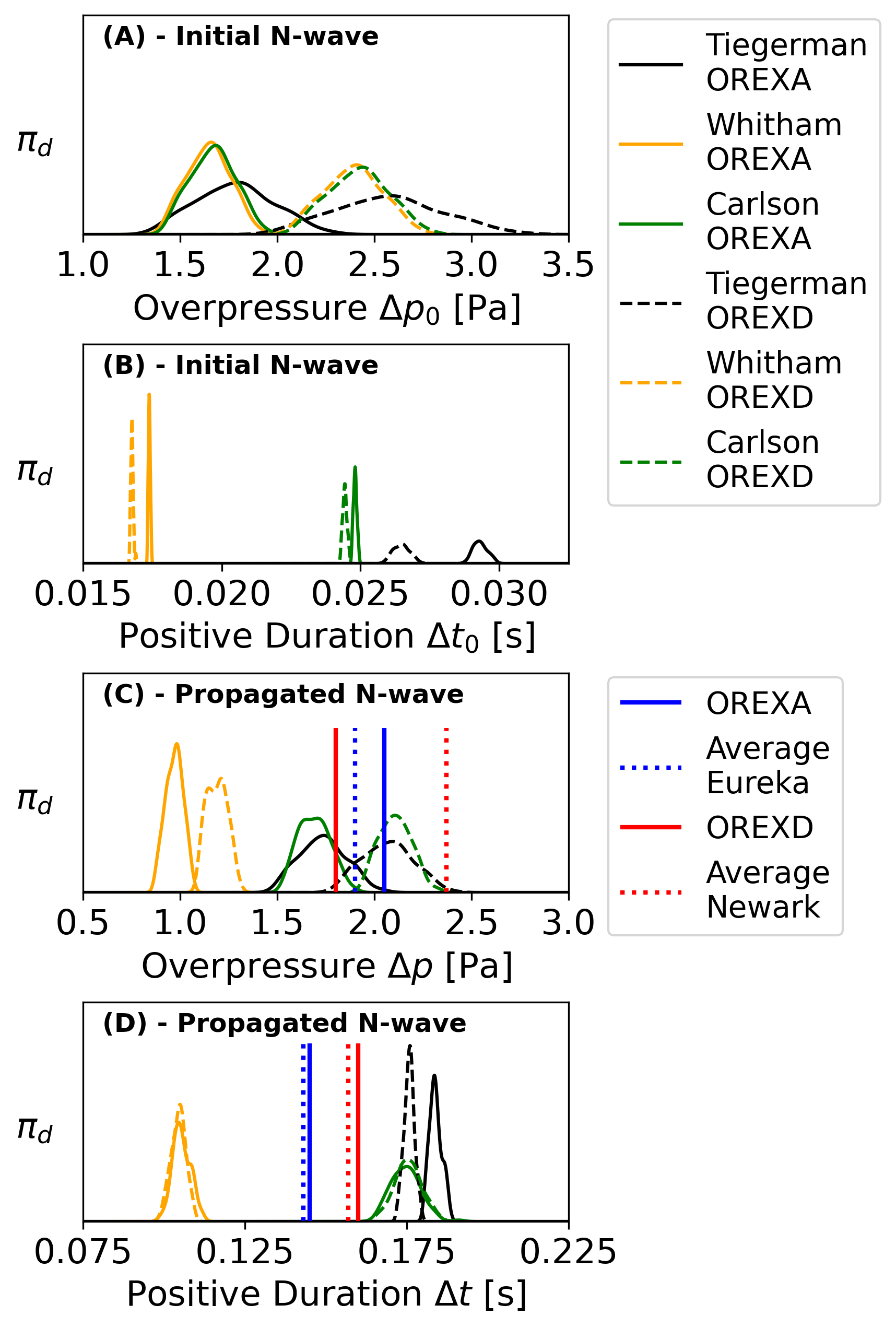}
    \caption{Probability density functions $\pi_d$ for (A) the initial N-wave overpressure $\Delta p_0$, (B) initial N-wave positive phase duration $\Delta t_0$, (C) predicted overpressure after propagation to the ground, and (D) predicted signal positive phase duration after propagation to the ground. The solid lines describe OREXA, and the dashed lines describe OREXD. The Whitham predictions are yellow, the Carlson predictions are green, and the Tiegerman predictions are blue. The recorded N-wave overpressure and positive phase duration for OREXA and OREXD are shown in (C) and (D) with the average values across the Eureka Airport stations and Newark Valley stations for reference (Table \ref{tbl:mics}).}
    \label{fig:PDFs2}
\end{figure}

\begin{figure}
    \centering
    \includegraphics[scale=0.7]{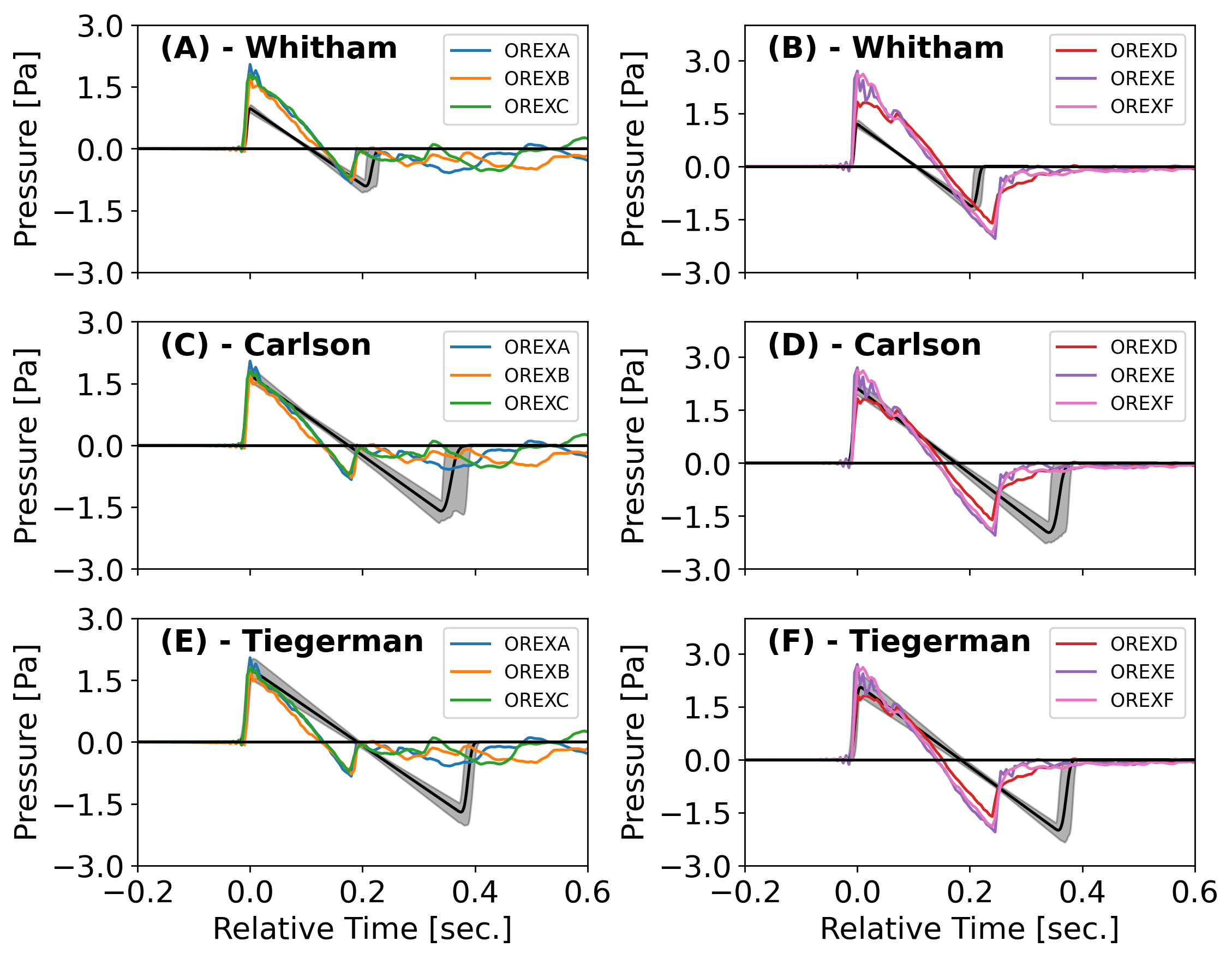}
    \caption{Comparison of time-aligned predicted waveforms with recorded waveforms. The black lines are the mean simulated waveforms, and the shaded regions denote the range of values from the propagation uncertainty - the minimum to maximum at each point in time. The left column shows the Eureka Airport stations and the right column shows the Newark Valley stations. (A) and (B) are the propagated N-waves with initial values from Whitham's Far Field Theory (Section \ref{section:Whitham}). (C) and (D) show propagated N-waves with initial values from Carlson's simplified sonic boom prediction method (Section \ref{section:Carlson}). (E) and (F) show propagated N-waves constructed from the Tiegerman's Drag Dominated hypersonic model (Section \ref{section:Tiegerman}). The initial N-wave parameter distributions and the distributions after propagation are summarized in Figure \ref{fig:PDFs2}.}
    \label{fig:N_wave_comparison}
\end{figure}

The propagated waveforms are shown with the recorded data and in Figures \ref{fig:N_wave_comparison}E and \ref{fig:N_wave_comparison}F. We overall have excellent agreement between the recorded data aligned to the peak overpressure and the propagated waveforms. The propagated waveforms have a larger rise time though, and the rarefaction lobe of the N-waves for the Eureka Airport stations is notably smaller than the predicted waveform. The N-waves recorded at Eureka Airport are unbalanced, and there is a coherent signal after the tail shock (Figure \ref{fig:fig2}). This suggests that the recorded tail shock may be truncated by the arrival of a second, un-modeled, signal due to the turbulent wake of the SRC. The distributions of predicted overpressure $\Delta p$ and positive phase duration $\Delta t$ of the propagated waveforms are shown in Figures \ref{fig:PDFs2}B and \ref{fig:PDFs2}C.

\section{Discussion}
Recordings from the primary boom carpet of the OSIRIS-REx SRC are used to evaluate a suite of source models, and uncertainty due to the propagation is incorporated into the predicted waveforms. In general, waveforms recorded in Newark Valley had larger amplitudes than those recorded at Eureka Airport (Figure \ref{fig:stns}; Table \ref{tbl:mics}), and all three overpressure prediction methods predict this trend (Figures \ref{fig:PDFs1}, \ref{fig:PDFs2}, and \ref{fig:N_wave_comparison}). All recorded waveforms have perturbations around the peak amplitude. The measured signal durations $\tau$ (Table \ref{tbl:mics}) are also somewhat less than twice the positive phase duration $\Delta t$, which would be expected for a balanced N-wave. This effects may be the result of propagation through turbulence in the atmosphere \citep{Crow1969, Luquet2015, Lonzaga2023, Iura2024}, and the superposition of a turbulent wake signal (Figure \ref{fig:fig2}) may also contribute to the apparent shortening of the recorded N-waves. While these phenomena need to be considered for more accurate modeling of the recorded signals, they are not currently accounted for in the analytical or numerical methods evaluated here.

The source locations for N-waves recorded at Eureka Airport and Newark Valley were determined using the predicted trajectory (Figure \ref{fig:stns}) and an ensemble of atmospheres (Figure \ref{fig:atmos}). For each atmosphere realization, a grid search was used to determine the location that minimized the distance between ray bounce points and the station locations (Figure \ref{fig:orex_loc}A; Table \ref{tbl:Locations}). For comparison, the Eureka Airport beamforming results in Figure \ref{fig:fig2} were used to estimate an incidence angle, and a geometric approach was used to estimate an altitude of 50 km, which underpredicts the altitude by approximately 10 km. This appears to be due to neglecting the refraction of acoustic rays near the ground, which would make the signal arrive at an apparently shallower incidence angle. While the as-flown trajectory is unknown, it appears that our calculations may be somewhat weakly sensitive to the differences (Figure \ref{fig:N_wave_comparison}). In cases where the source trajectory is completely unknown, a procedure of trajectory estimation \citep[e.g.,][]{Nishikawa2022} and then waveform prediction may be used.

The signals recorded at Eureka Airport were emitted from higher altitudes than the signals recorded at Newark Valley (Table \ref{tbl:Locations}), and the recorded signals on the ground at Eureka Airport generally have lower amplitudes than the signals recorded at Newark Valley. This suggests that the higher source amplitude resulted in a lower signal amplitude \citep{Silber2025}. Previous work investigating the effect of altitude on acoustic-gravity wave amplitudes from spherical explosions in the atmosphere suggests that signal strength decreases with altitude above approximately 50 km for extremely large, low-frequency sources \citep{Pierce1971}. However, for higher frequency, sonic boom sources such as the Apollo ascents and the OSIRIS-Rex SRC re-entry, this decrease with altitude may be nearly monotonic \citep{Maglieri2011_Shuttle}.

The first source model we evaluated was the sonic boom prediction method of \citet{Whitham1952}. Overpressure predictions were 0.456 $\pm$ 0.030 Pa for OREXA and 0.553 $\pm$ 0.035 Pa for OREXD (Figures \ref{fig:PDFs1}E). These values are approximately one-forth to one-fifth the observed overpressures (Table \ref{tbl:mics}). The predicted signal positive phase durations are 0.114 $\pm$ 0.001 seconds for OREXA and 0.109 $\pm$ 0.001 seconds for OREXD (Figure \ref{fig:PDFs1}G). The positive phase durations predicted by Equations \ref{eqn:Whitham_eqns} are too short by approximately 0.02 - 0.03 seconds for the Eureka Airport stations and 0.03 - 0.04 seconds for the Newark Valley stations (Table \ref{tbl:mics}). A large disparity between the recorded and observed overpressures for Equations \ref{eqn:Whitham_eqns}, with a much smaller disparity in the signal duration, has been noted in other work \citep{Fernando2025}.

When used as a source for the Burgers' equation, the overpressures are 0.971 $\pm$ 0.047 Pa for OREXA and 1.190 $\pm$ 0.055 Pa for OREXD (Figures \ref{fig:PDFs2} and \ref{fig:N_wave_comparison}). The corresponding signal positive phase duration predictions are 0.105 $\pm$ 0.003 s (OREXA) and 0.104 $\pm$ 0.002 s (OREXD). Numerical propagation through the ensemble of atmospheres nearly doubled the predicted overpressures compared to the analytic model, but the predicted signals are also shorter. While the Burgers' equation propagation somewhat improved the overpressure predictions of the Whitham model, a relatively large disparity remains between the this model's predictions and the recorded data (Figures \ref{fig:PDFs2} and \ref{fig:N_wave_comparison}).

Whitham's analytic F-function approximation to the flow fields begins to break down above approximately Mach 3 \citep{Whitham1952, Carlson1972, Gottlieb1988}. This is due to the the changing flow conditions around the body as the projectile approaches the hypersonic flow regime \citep{Anderson2006}. Additionally, as a first-order correction to the linearized theory, it holds for slender projectiles. This means that the slope of the projectile is less than the Mach angle $\alpha$ \citep[e.g.,][]{Gottlieb1988}. The approximately 60$^\circ$ angle for the SRC is much greater that the Mach angle for the 10s of Mach examined in this manuscript, which means that the SRC is a blunt body. An interpolating polynomial method \citep{Ritzel1988} was used to evaluate the F-function (Equation \ref{eqn:F_function}) for an approximate model of the SRC (Figure \ref{fig:Carlson}). However, the resulting F-function was unbalanced, which resulted in our use of Equation \ref{eqn:K_delta} instead of Equation \ref{eqn:K_W} to calculate the shape factor K. While they may be a poor approximation at the Mach numbers examined here, Equations \ref{eqn:Whitham_eqns} are analytic and well known, so they are a useful reference point for the other source models.

The Carlson sonic boom prediction method \citep{Carlson1978} was developed by adding simulation-derived factors to the Whitham model to correct for a nonuniform atmosphere. This approach was used to predict approximate overpressure values for the Apollo 15 and 16 command module re-entries \citep{Carlson1978}, which suggests that this method can better accommodate a blunt body than the original Whitham method \citep{Whitham1952}. The predicted OSIRIS-REx re-entry overpressure values are 1.859 $\pm$ 0.123 Pa for OREXA and 2.228 $\pm$ 0.140 Pa for OREXD (Figure \ref{fig:PDFs1}F). The predicted positive phase duration values are 0.102 $\pm$ 0.001 seconds for OREXA and 0.100 $\pm$ 0.001 seconds for OREXD (Figure \ref{fig:PDFs1}H). These overpressure predictions are much closer to the recorded data (Table \ref{tbl:mics}) than those from Equations \ref{eqn:Whitham_eqns}. The mean overpressure values across the Eureka Airport stations (1.90 Pa) and the Newark Valley stations (2.37 Pa) are within or near the one standard deviation ranges (Figure \ref{fig:PDFs1}F). However, the positive phase durations again appear shorter than the recorded data (Figure \ref{fig:PDFs1}H). This difference is now approximately 0.03 - 0.05 seconds for the Eureka Airport stations and 0.05 - 0.06 seconds for the Newark Valley stations.

When the Carlson model is used as a source model for propagation with the Burgers' equation, the predicted overpressures for this method are 1.692 $\pm$ 0.086 Pa for OREXA and 2.108 $\pm$ 0.088 Pa for OREXD. The corresponding signal positive phase duration predictions are 0.175 $\pm$ 0.010 s (OREXA) and 0.175 $\pm$ 0.004 s (OREXD). The numerical predictions for overpressure with the Carlson model show a smaller difference compared to Equations \ref{eqn:Carlson_eqns} than the corresponding predictions with the Whitham source model, but these predicted amplitudes are smaller. However, the predicted signal durations are longer (Figures \ref{fig:PDFs2} and \ref{fig:N_wave_comparison}).

As opposed to the other source models, which are shape factor methods, the source model for drag-dominated vehicles \citep{Tiegerman1975} models the shock waves from the SRC as radiating from a point explosion with cylindrical symmetry. Acoustic source models for lightning \citep{Jones1968} and bolides \citep{ReVelle1974, Edwards2009, Silber2019} have a similar initial framework. For drag dominated vehicles, Equations \ref{eqn:Tiegermanp} and \ref{eqn:Tiegermant} are used as boundary conditions for an inviscid Burgers' equation. This approach predicted overpressures of 1.736 $\pm$ 0.126 Pa for OREXA and 2.108 $\pm$ 0.088 Pa for OREXD. The signal durations were predicted to be 0.184 $\pm$ 0.002 seconds for OREXA and 0.176 $\pm$ 0.002 seconds for OREXD (Figures \ref{fig:PDFs2} and \ref{fig:N_wave_comparison}). These overpressure predictions are again near the mean overpressure values across the Eureka Airport stations (1.90 Pa) and the Newark Valley stations (2.37 Pa), but the signal positive phase durations are longer than the data by 0.03 - 0.05 seconds for OREXA and 0.02 seconds for OREXD (Figure \ref{fig:PDFs1}F).

High fidelity numerical modeling has shown that certain assumptions in the Carlson and Tiegerman models fail to capture effects that may be visible on the propagation scales considered in this work. These more accurate physical models may explain some of the differences between the predicted and recorded waveforms (Figure \ref{fig:N_wave_comparison}). First, the Tiegerman model is a line source model, and simulation results show that a finite source size has lower overpressure than a corresponding line source \citep{Henneton2015}. As the scaled nonlinearity $\tilde{\beta}$ is proportional to the reference pressure (Equation \ref{eqn:burgers2}), this suggests that infinite line sources may overpredict signal duration. Additionally, real gas equations of state and viscous effects from the boundary layer are also not considered in these models, which assume an ideal gas and coupling to an inviscid Burgers' equation. Real gas effects lower predicted overpressure and increase duration relative to an ideal gas assumption \citep{Henneton2015, King2024}, and viscous effects increase overpressure and signal duration relative to inviscid flow \citep{King2024}. Finally, we note that hypersonic sources with large Mach numbers, such as meteorites, may more closely resemble blast waves than symmetric N-waves in the far field \citep{Henneton2015, Yamashita2016, Nemec2017}. We are particularly interested to further investigate these ``caret-waves" in future work.

Here we examine a suite of semi-analytic source models for the OSIRIS-REx SRC re-entry at hypersonic speeds. Both applications of the Carlson model \citep{Carlson1978} and the propagation of the Tiegerman model \citep{Tiegerman1975} yield similar, somewhat accurate, predictions for overpressure that overlap in range when considering uncertainty in the propagation medium. We note that bolide prediction models appear to share characteristics with both models \citep{ReVelle1974, Edwards2009, Silber2019} - propagation correction factors as well the cylindrical point explosion framework. Further examination of these connections between models as well as the propagation to farther boom carpets is the topic of future work.

\section{Conclusions}
We use six N-wave recordings of the September 24th, 2023 OSIRIS-REx sample return capsule hypersonic re-entry \citep{Lauretta2017} to evaluate a suite of sonic boom source models with uncertainty in the propagation medium. The OSIRIS-REx source dimensions are well characterized, and an approximate trajectory is known, which is somewhat rare for infrasonic sources. With an ensemble of atmospheric states \citep{Blom2023}, a Mach cone prediction capability \citep{Blom2024_MachCone}, and a Burgers' equation solver \citep{Lonzaga2015, Blom2021_Thermosphere} that are all provided as open source software, we predicted N-waves from the OSIRIS-REx SRC. The simplified sonic boom prediction method \citep{Carlson1978} and drag dominated source model \citep{Tiegerman1975} produced more accurate predictions than the original Whitham model \citep{Whitham1952}. While predicted overpressures are a good match for the recorded amplitudes, the signal duration was somewhat overpredicted (Figure \ref{fig:N_wave_comparison}).

\section{ACKNOWLEDGEMENTS}
We thank the two anonymous reviewers for comments that greatly improved the quality of this manuscript. This research was supported by Los Alamos National Laboratory Laboratory Directed Research and Development through project numbers 20230120DR and 2020188DR as well as through its Center for Space and Earth Science (CSES). CSES is funded by LANL’s Laboratory Directed Research and Development (LDRD) program under project number 20240477CR. The authors thank Elizabeth Silber, Joel Lonzaga, and Daniella Mendoza DellaGiustina for helpful comments at different stages of this work and Carly M. Donahue, Lo\"{i}c Viens, Luke B. Beardslee, Elisa A. McGhee, and Lisa R. Danielson for support installing the infrasound microphones. This work was supported by the U.S. Department of Energy through the Los Alamos National Laboratory. Los Alamos National Laboratory is operated by Triad National Security, LLC, for the National Nuclear Security Administration of U.S. Department of Energy (Contract No. 89233218CNA000001).

\FloatBarrier
\section{APPENDIX}
Due the similar location of the source for the stations at Eureka Airport and at Newark Valley, OREXA and OREXD were taken as representative stations for comparison and discussion. In Table \ref{tbl:All_predictions} we provide a summary of all the predictions from each method in Section IV. The ``Whitham Analytic" notation refers to using Equations \ref{eqn:Whitham_eqns} to predict N-wave overpressure and signal positive phase duration on the ground. The notation ``Carlson Analytic" refers to using Equations \ref{eqn:Carlson_eqns} to predict N-wave overpressure and signal duration on the ground. The ``Burgers' Equation'' notation refers to using the associated source models to generate an initial N-wave, which is the numerically propagated to the ground with Equation \ref{eqn:Burgers_eqn}.

\begin{table}
\begin{center}
    \caption{All overpressure $\Delta p$ [Pa] and positive phase duration $\Delta t$ [s] predictions.}
    \label{tbl:All_predictions}
    \begin{tabular}{ c c c}
    Method & OREXA & OREXD \\
    \hline
     Whitham Analytic $\Delta p$ [Pa] & 0.456 $\pm$ 0.030 & 0.553 $\pm$ 0.035 \\
     Whitham Analytic N-wave $\Delta t$ [s] & 0.114 $\pm$ 0.001 & 0.109 $\pm$ 0.001 \\
     Carlson Analytic $\Delta p$ [Pa] & 1.859 $\pm$ 0.123 & 2.228 $\pm$ 0.140 \\
     Carlson Analytic N-wave $\Delta t$ [s] & 0.102 $\pm$ 0.001 & 0.100 $\pm$ 0.001 \\
     Whitham Burgers' Equation $\Delta p$ [Pa] & 0.971 $\pm$ 0.047 & 1.19 $\pm$ 0.055 \\
     Whitham Burgers' Equation N-wave $\Delta t$ [s] & 0.105 $\pm$ 0.003 & 0.104 $\pm$ 0.002 \\
     Carlson Burgers' Equation $\Delta p$ [Pa] & 1.692 $\pm$ 0.086 & 2.108 $\pm$ 0.088 \\
     Carlson Burgers' Equation N-wave $\Delta t$ [s] & 0.175 $\pm$ 0.010 & 0.175 $\pm$ 0.004 \\
     Tiegerman Burgers' Equation $\Delta p$ [Pa] & 1.736 $\pm$ 0.126 & 2.069 $\pm$ 0.136 \\
     Tiegerman Burgers' Equation N-wave $\Delta t$ [s] & 0.184 $\pm$ 0.002 & 0.176 $\pm$ 0.002 \\
    \end{tabular}
\end{center}
\end{table}

\section*{AUTHOR DECLARATIONS}
\subsection*{Conflict of Interest}
The authors declare that there are no conflicts of interest to disclose.

\section*{DATA AVAILABILITY}
The infrasound array processing tools used here are publicly available at \url{https://github.com/uafgeotools/lts_array} and \url{https://github.com/uafgeotools/array_processing}. The Mach cone modeling capability and the Burgers' equation solver are part of the \textit{InfraGA/GeoAc} software that is available at \url{https://github.com/LANL-Seismoacoustics/infraGA}. The atmospheric profile was constructed using the NCPA G2S Request System (\url{https://g2s.ncpa.olemiss.edu}) and ensembles were generated with the \textit{stochprop} software that is available at \url{https://github.com/LANL-Seismoacoustics/stochprop}.


\end{document}